\documentclass[twocolumn]{aastex631}
\usepackage{textcomp}
\usepackage{amssymb}
\usepackage{rotating}

\newcommand{\fermi}{\textit{Fermi}}
\newcommand{\gr}{$\gamma$-ray}
\newcommand{\rcw}{RCW~103}
\newcommand{\hess}{HESS J1616$-$508}

\begin{document}

\title{Revisiting Gamma-Ray Emission of the Supernova Remnant RCW 103}

\author{Yi Xing}

\affiliation{Key Laboratory for Research in Galaxies and Cosmology, Shanghai Astronomical Observatory, Chinese Academy of Sciences,
80 Nandan Road, Shanghai 200030, China; yixing@shao.ac.cn; wangzx20@ynu.edu.cn}

%%\author{Zhongxiang Wang}
\author[0000-0003-1984-3852]{Zhongxiang Wang}
\affiliation{Department of Astronomy, School of Physics and Astronomy,
Yunnan University, Kunming 650091, China}
\affiliation{Key Laboratory for Research in Galaxies and Cosmology,
Shanghai Astronomical Observatory, Chinese Academy of Sciences,
80 Nandan Road, Shanghai 200030, China; yixing@shao.ac.cn; wangzx20@ynu.edu.cn}

\author{Dong Zheng}
\affiliation{Department of Astronomy, School of Physics and Astronomy,
Yunnan University, Kunming 650091, China}

\begin{abstract}
We analyze more than 15 years of the \gr\ data, obtained with the Large Area
	Telescope (LAT) onboard {\it the Fermi Gamma-ray Space Telescope 
	(Fermi)}, for the region of the young supernova remnant (SNR) RCW~103,
	since the nearby source 4FGL J1616.2$-$5054e, counterpart to 
	HESS~J1616$-$518 and $\simeq$13\,arcmin away from the SNR, is 
	determined to be extended 
	in the more recent \fermi-LAT source catalog. Different templates
	for 4FGL J1616.2$-$5054e and RCW~103 are tested, and we find that a
	point source with a power-law (PL) spectrum at the southern limb of 
	the SNR best describes the detected \gr\ emission. 
	The photon index of the PL emission is $\Gamma\simeq 2.31$ (or 
	$\alpha\simeq 2.4$ in a Log-Parabola model), 
	softer than the previously reported 
	$\Gamma\simeq 2.0$ when the counterpart to HESS~J1616$-$518 was 
	considered to be a point source (which likely caused mis-identification
	of extended emission at RCW~103). 
	In order to produce the \gr\ emission in a hadronic scenario,
	we estimate that protons with an index$\sim$2.4 PL energy distribution 
	are needed. These results fit with those from multi-wavelength 
	observations that have indicated the remnant
	at the southern limb is interacting with a molecular cloud.
\end{abstract}
\keywords{Gamma-rays (637); Supernova remnants (1667)}

\section{Introductions}

\begin{figure*}
	\centering
   \includegraphics[width=0.47\textwidth]{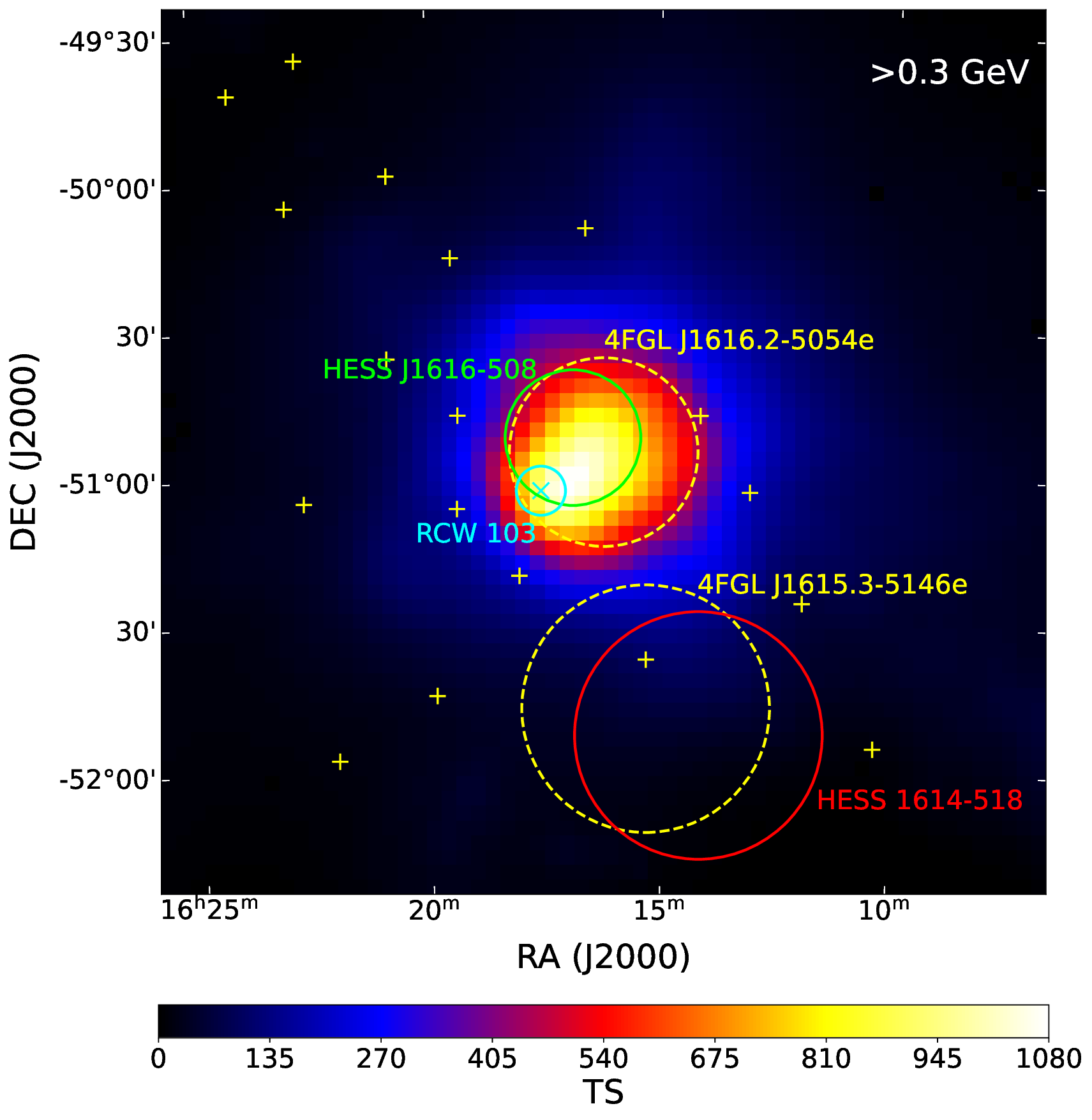}
   \includegraphics[width=0.47\textwidth]{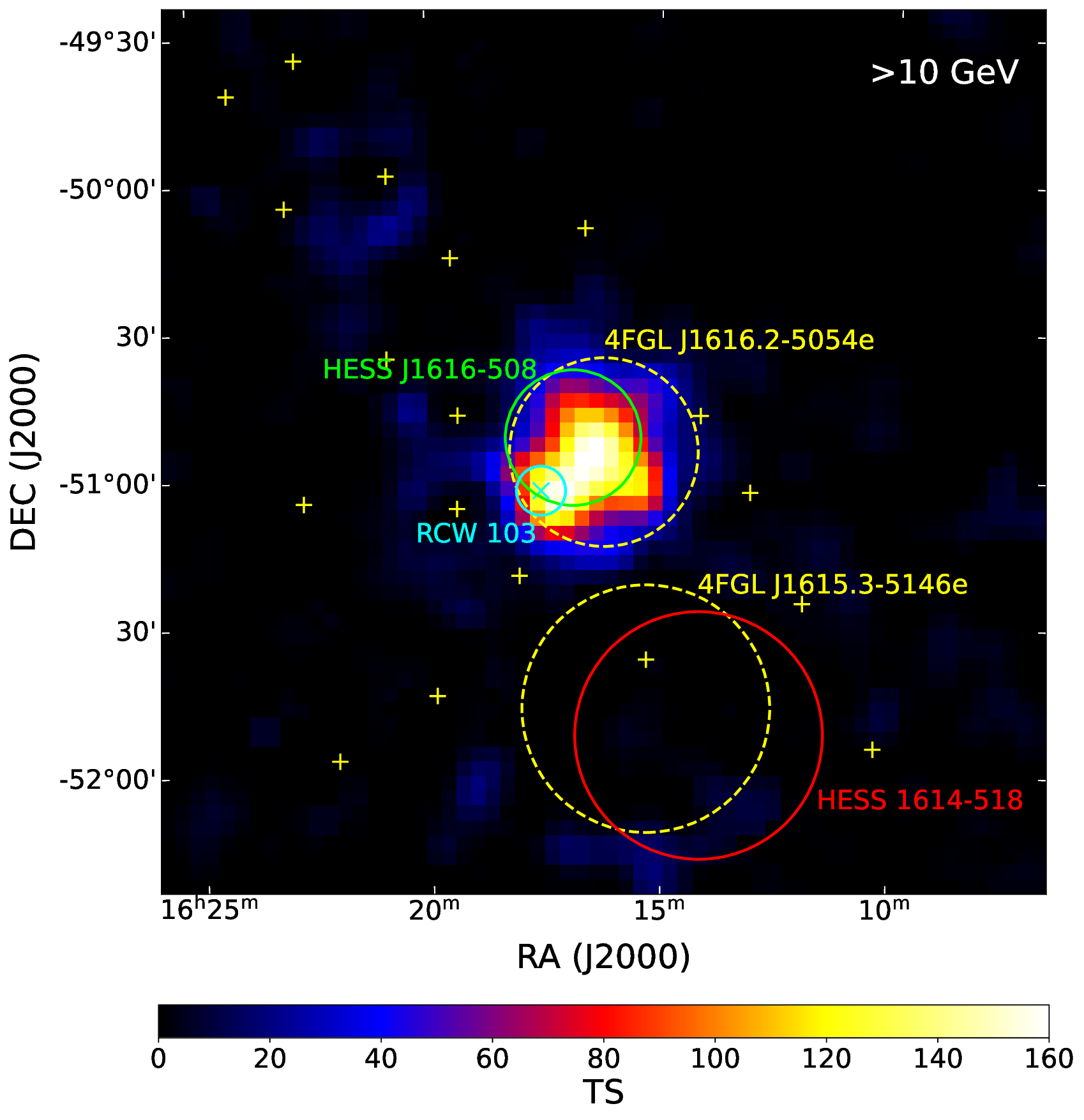}
   \includegraphics[width=0.37\textwidth]{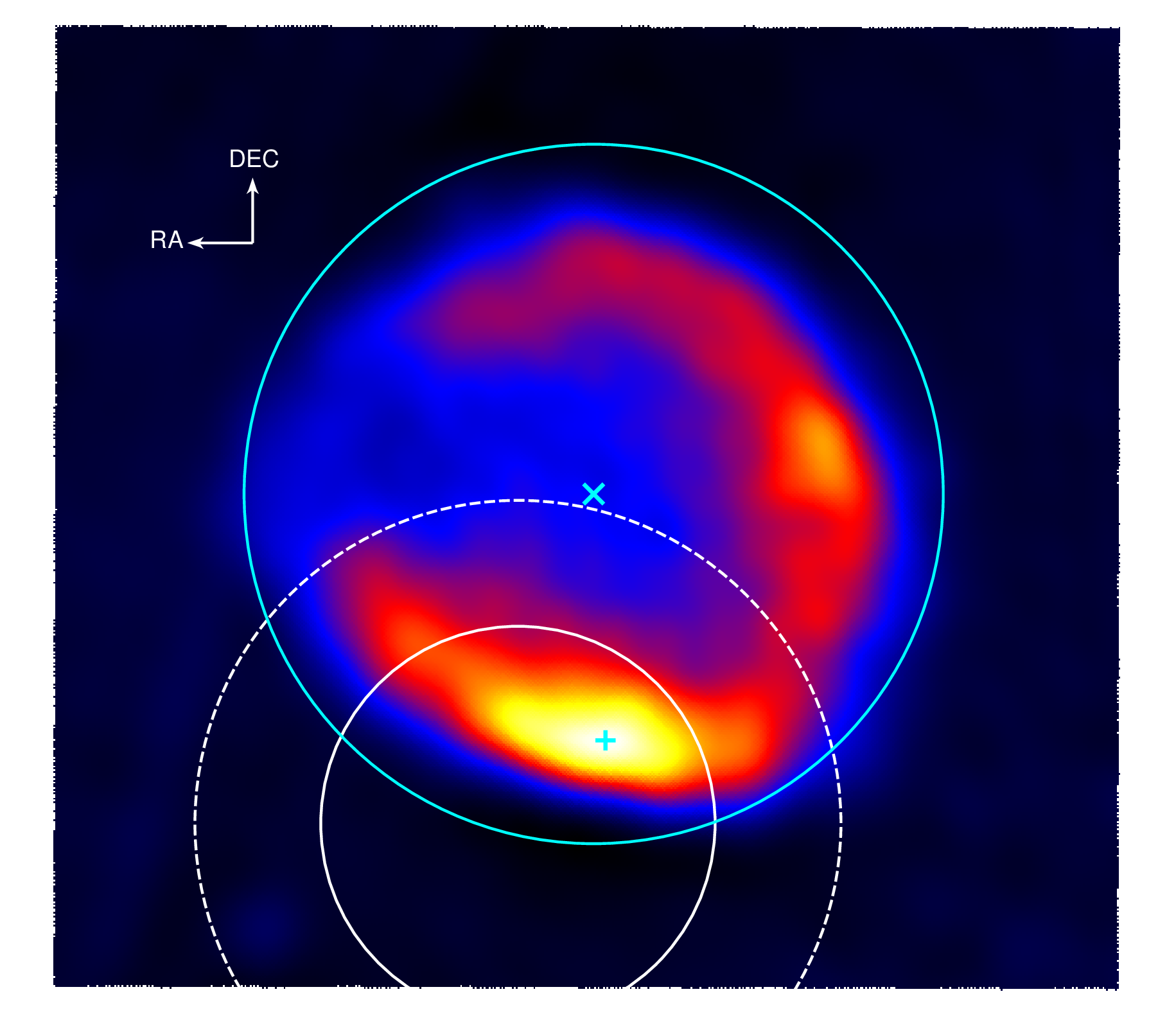}
	\caption{{\it Top}: TS maps for the region of \hess\ and \rcw\ in 
	the energy ranges of $\geq$0.3\,GeV and $\geq 10$\,GeV ({\it left} 
	and {\it right} panel, respectively). The known \fermi-LAT sources in
	the field (marked with plus signs) are removed (i.e., included 
	in
	the source model and thus not shown in the TS maps), and the extended
	source 4FGL~J1615.3$-$5146e (marked with a dashed circle) is also 
	removed. The solid circles mark the regions of the VHE sources 
	\hess\ and HESS J1614$-$518. The region of \rcw\ and its CCO are marked
	with a small cyan circle and a cyan cross, respectively. 
	{\it Bottom}: MOST radio map of the region. The same symbols as in
	top panels are used to mark the region of \rcw\ and its CCO, while
	the cyan plus marks the brightest position in the radio map.
	The white solid and dashed circles mark the 1$\sigma$ and 2$\sigma$ 
	error circles of the PS source in \rcw\ detected in \fermi-LAT data. 
}
   \label{fig:tsmap}
\end{figure*}

The supernova remnant (SNR) RCW~103 (or G332.4$-$0.4; \citealt{rcw60}) has been
extensively studied at multi-wavelengths. Its role as a valuable 
sample of a young SNR, as well as its product, 1E~161348$-$5055 as
a likely 
magnetar (or previously known as a central compact object; 
\citealt{tg80, dai+16, rbe+16}), have attracted much attention to it.
Adopting its most recently derived distance of 3.1\,kpc \citep{rgj+04}, 
its age has been estimated to be $\sim$2\,kyr \citep{cdb97} or $\sim$4\,kyr
\citep{bsf19,lrg20}. The SNR appears in the sky as a nearly circular, 
shell-like structure, with a diameter of 10\arcmin. Notably, its southern
part is brighter than the regions of other directions 
(Figure~\ref{fig:tsmap}), which has been seen
at multi-wavelengths from radio frequencies to X-rays (e.g., \citealt{pin+11}).

SNRs are primary \gr\ sources in the Galaxy (e.g., \citealt{zyl19,4fgl-dr4}).
Cosmic-ray particles are produced from their shock fronts through the
diffuse shock acceleration process (e.g., \citealt{dru83}). The particles
emit $\gamma$-rays via the hadronic or/and leptonic processes; the
former involves proton-proton collisions and the latter electron radiation
processes including synchrotron and inverse Compton scattering (ICS).

Previously, \citet{xing+14} has reported likely detection of extended
\gr\ emission from \rcw\ in the 1--300\,GeV data obtained with the Large 
Area Telescope (LAT) on board the {\it Fermi Gamma-ray Space Telescope
(Fermi)}. At the time, only 5 years of the \fermi-LAT data were collected
and the source model for the data analysis was based on the \fermi-LAT second 
source 
catalog \citep{nol+12}. In addition, the SNR is located in a crowded region,
with two very-high-energy (VHE) sources nearby. They are HESS~J1616$-$508
and HESS~J1614$-$518, which were detected in the High Energy
Stereoscopic System (HESS) survey \citep{hess05,hess06,hgps}.
In particular, the first source is $\sim$13\,arcmin away from \rcw, considered
to be very close given the spatial resolutions of \fermi-LAT and VHE detectors.
In the previous analysis, the GeV counterpart to \hess\ 
was taken as a point source (2FGL J1615.0$-$5051), but according to
the latest \fermi\  LAT source catalog (4FGL-DR4; \citealt{4fgl-dr4}), 
both GeV counterparts
to \hess\ and HESS~J1614$-$518 are extended sources, named as 
4FGL~J1616.2$-$5054e
and 4FGL~J1615.3$-$5146e respectively (see Figure~\ref{fig:tsmap}).
With the updated source information taken into consideration,
we re-analyzed $>$15 years of the \fermi-LAT data for \rcw,
and obtained new results.
In this paper, we report the results.

\section{Data Analysis and Results}
\label{sec:da}

\subsection{\textit{FERMI}-LAT data and source model}
\label{sec:obs}

We selected the LAT events in energy range of from 0.1 to 500 GeV 
from the updated \textit{Fermi} LAT Pass 8 database (P8R3).
The region of interest (RoI) was set to be centered at \hess\ with a size
of 20$^{\circ}$ $\times$ 20$^{\circ}$.
The start and end time of the events were August 4 2008, 15:43:39 (UTC) 
and December 11 2023, 23:26:35 (UTC), respectively.
We included the events in the \texttt{SOURCE} event class with zenith angles 
less than 90\,degree to avoid the Earth's limb contamination, and
excluded events with `bad' quality flags.
Both selections are recommended by the LAT 
team\footnote{\footnotesize http://fermi.gsfc.nasa.gov/ssc/data/analysis/scitools/}.

A source model was constructed by including all sources within a 20-degree 
radius of \hess\ based on 4FGL-DR4 \citep{4fgl-dr4}. The positions and
spectral parameters for these sources provided in the catalog were adopted. 
For sources within 5-degree radius of \hess, we set their spectral 
parameters free, and for the remaining sources, their spectral parameters
were fixed at their catalog values. In this latest source catalog, 
the counterpart of \hess\ is an extended source described by a uniform disk 
with a radius of 0\fdg32, and its \gr\ spectrum is modeled with a
power law (PL).
The spectral model gll\_iem\_v07.fit and spectral file 
iso\_P8R3\_SOURCE\_V3\_v1.txt were included as the Galactic and 
extragalactic diffuse emission, respectively. The normalizations of 
the two diffuse-emission components were always set as free parameters 
in our analysis.

In constructing 4FGL catalogs \citep{4fgl}, the LAT events below 316\,MeV 
were given low weights, especially in the vicinity of the Galactic plane.
In addition, only \texttt{Front} events were utilized in the low energy 
band below 316\,MeV \citep{4fgl} in order to avoid the relatively poor point 
spread function (PSF) and the contamination from the Earth limb. Given these, 
we did not include the events below 300\,MeV in our analysis since the target 
is located in the Galactic plane with a few nearby sources around.

\subsection{Maximum Likelihood Analysis}
\label{sec:mla}

With the source model, we performed the standard binned likelihood analysis
to the LAT data in 0.3--500\,GeV. A PL photon index $\Gamma$=2.02$\pm$0.03 
and a 0.3--500\,GeV photon flux 
$F_{0.3-500}\simeq (3.5\pm 0.2)\times 10^{-8}$\,photon\,cm$^{-2}$\,s$^{-1}$ 
were obtained for \hess, and the Test Statistic (TS) value was 1640. 

To examine the source field, we calculated TS maps centered at \hess\ 
in different energy ranges. All sources in the source model, except 
the counterpart of \hess, were included and thus removed in the TS maps. 
In Figure~\ref{fig:tsmap}, we showed TS maps in the energy ranges of 
0.3--500\,GeV and 10--500\,GeV as the representative ones, where the latter 
helps reveal the emission in a high, $\lesssim 0\fdg2$ spatial resolution\footnote{\footnotesize https://www.slac.stanford.edu/exp/glast/groups/canda\\/lat\_Performance.htm} 
(smaller than the separation between \hess\ and \rcw). 
As can be seen in the low-energy TS map, possibly extended emission within 
the source region of \hess\ \citep{hgps} is detected, which is 
enclosed by the extended source 4FGL~J1616.2$-$5054e given in 4FGL-DR4. 
We also noted that the other 
extended source 4FGL~J1615.3$-$5146e, corresponding to HESS~J1614$-$518, is
cleanly removed. However, the position of the highest TS values within
4FGL~J1616.2$-$5054e is off the center towards the south-east direction, where
\rcw\ is located.
In the high-energy TS map, the high TS-value region is possibly composed of
two components, one slightly off the center of 4FGL~J1616.2$-$5054e and
the other close to the location of \rcw. Thus, it is likely that emission
from \rcw\ was detected even though it is close to an extended source.
\begin{table}
\begin{center}
\caption{Comparison of maximum likelihood values}
\label{tab:lh}
\begin{tabular}{lcccccccc}
\hline
Source Model & 2$\times (\log L-\log L_{0}$) \\ \hline
HESS$_{0.32}$ & 0 \\
HESS$_{0.40}$ & 43.2 \\
	PS $+$ HESS$_{0.32}$ & 57.1 \\
Radio$_{\rm disk}$ $+$ HESS$_{0.32}$ & 63.5 \\
Radio$_{\rm profile}$ $+$ HESS$_{0.32}$ & 71.3 \\
Radio$_{\rm ps}$ $+$ HESS$_{0.32}$ & 82.1 \\
Radio$_{\rm ps}$ $+$ HESS$_{0.38}$ & 106.4 \\
\hline
\end{tabular}
\\
\footnotesize{}
\end{center}
\end{table}

\subsection{Spatial Distribution Analysis and Source Identification}
\label{sec:ss}

In order to fully probe the \gr\ emission in the region, we ran a series of
spatial analysis with different spatial templates considered. The base
model was that given in 4FGL-DR4, which sets a 0\fdg32 uniform disk
template for the GeV emission of \hess\ (HESS$_{0.32}$ model). All the  
likelihood results were compared to the likelihood $L_0$ of this base model.

We first tested a uniform disk with radius ranging from 0\fdg02 to 0\fdg50 
(with a step of 0\fdg02) centered at \hess\ as a check for the source's
GeV \gr\ emission. A simple PL was used to model the emission. 
We found that when radius was 0\fdg40 (HESS$_{0.40}$ model), 
a maximum $\log L$ value was obtained, which indicates a fit improvement 
at a 6.6$\sigma$ significance level (calculated from 
$\sqrt{2\times (\log L- \log L_0)}$; Table~\ref{tab:lh}) compared to 
the HESS$_{0.32}$ model. 

\begin{figure*}
	\centering
   \includegraphics[width=0.47\textwidth]{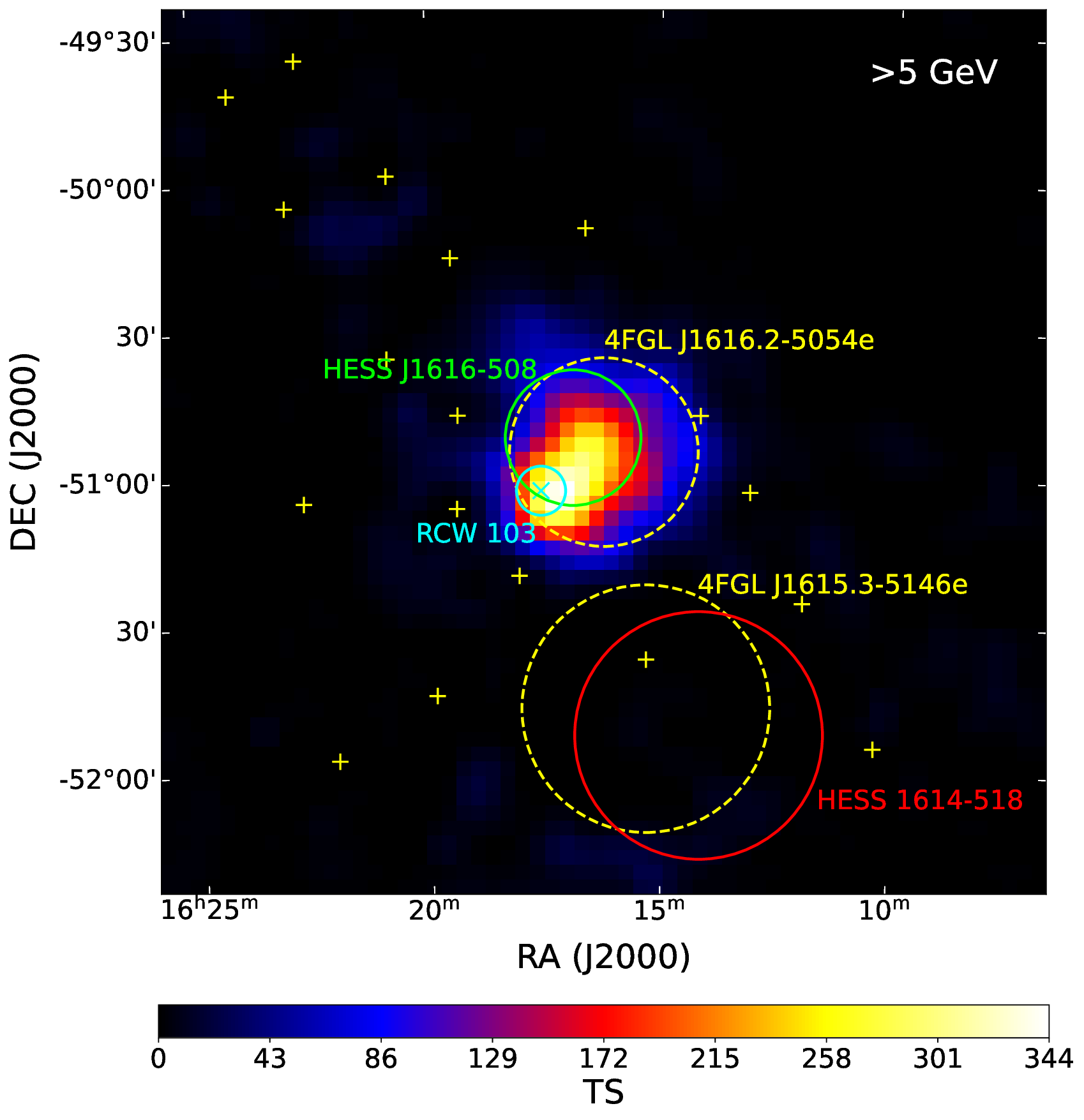}
   \includegraphics[width=0.47\textwidth]{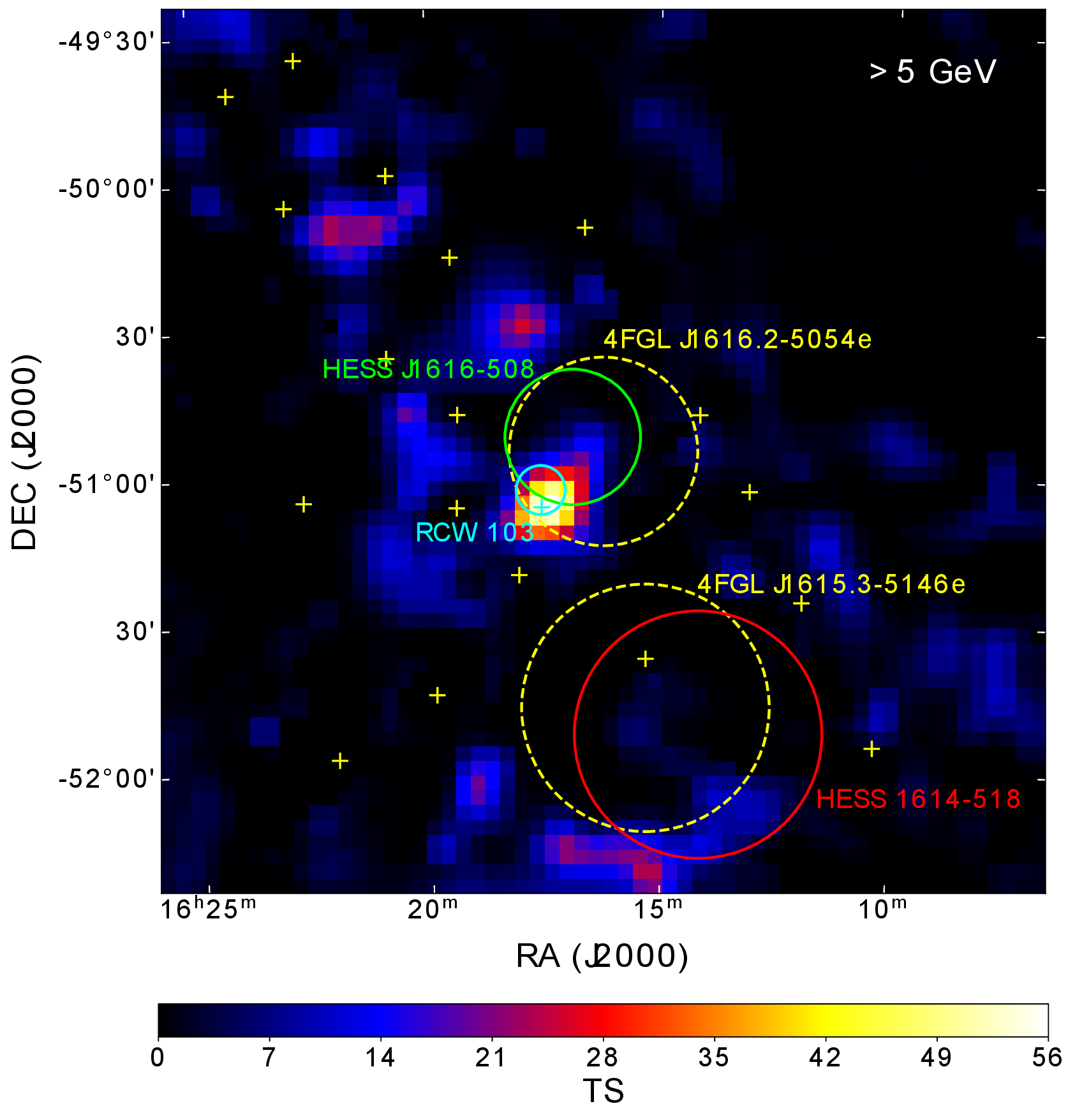}
	\caption{TS maps for the region of \hess\ and \rcw\ in the energy range
	of 5--500\,GeV, with the GeV counterpart to \hess\ kept ({\it left})
	and removed ({\it right}). The symbols are the same as in 
	Figure~\ref{fig:tsmap}. }
   \label{fig:ts5gev}
\end{figure*}

We then added different model templates in the source model at the position
of the central compact object (CCO) of \rcw. The templates included 
a point source (PS; PS model), a uniform disk with a radius of
5$\arcmin$ (given \rcw's size of 10$\arcmin$ in diameter detected at 
radio frequencies; Radio$_{\rm disk}$ model), and an emission template 
derived from 
the Molonglo Observatory Synthesis Telescope (MOST) radio map\footnote{\footnotesize http://snrcat.physics.umanitoba.ca/\\SNRrecord.php?id=G332.4m00.4} 
(Radio$_{\rm profile}$ model).  For \hess, the catalog
HESS$_{0.32}$ template was still used. Performing the maximum likelihood 
analysis to the data with each of the models, PS $+$ HESS$_{0.32}$,
Radio$_{\rm disk}$ $+$ HESS$_{0.32}$, and 
Radio$_{\rm profile}$ $+$ HESS$_{0.32}$,
the likelihood values were obtained and compared to that of the base model
(see Table~\ref{tab:lh}). The model of Radio$_{\rm profile}$ $+$ HESS$_{0.32}$ 
provided the best fit,
as it induced a 8.4$\sigma$ fit improvement compared to only HESS$_{0.32}$.
In addition, we noted that the fit improvement was 5.3$\sigma$ when
the non-catalog model HESS$_{0.40}$ we obtained above was considered for 
comparison.
These results indicate that extra emission is present at the position
of \rcw.

We further removed the counterpart of \hess\ from TS maps
to reveal the residual emission in the \rcw\ region. Two models
for the \hess\ source were considered, one the HESS$_{0.32}$ model and 
the other a uniform disk with a radius of 0\fdg38 (HESS$_{0.38}$ model).
The latter was used 
because of the results obtained in the following analysis. As a check,
we first calculated two TS maps in 0.3--5\,GeV respectively with the two
models and 
verified that no significant differences between the two models were found.
We then calculated TS maps in 5--500\,GeV without and with
HESS$_{0.38}$ being removed (Figure~\ref{fig:ts5gev}). Comparing the
two TS maps, it is clear to see the removal of the emission from \hess\ and
the presence of the emission at \rcw. However, the latter did not exactly
match the SNR's circular region, suggesting further analysis investigation 
would be needed.

We ran \texttt{gtfindsrc} in the {\tt Fermitools} to the $>$5 GeV data and
determined the position of the residual emission. We obtained
R.A.=244\fdg427, Decl.=$-$51\fdg119 (equinox J2000.0) with a 1$\sigma$ nominal 
uncertainty of 0\fdg047. The CCO is 0\fdg081 away from this position and 
outside of the 2$\sigma$ error circle (0\fdg077). Matching the position
and 1$\sigma$/2$\sigma$ error circles to the MOST radio map of \rcw\ 
(Figure~\ref{fig:tsmap}), it can be seen that the source region is actually
over the brightest part of the SNR's radio shell. 
Given this, we set a model for the residual emission which is at the position 
of the brightest part of the shell (Radio$_{\rm ps}$ model), and performed 
the likelihood analysis. The used \gr\ template for \hess\ was HESS$_{0.32}$. 
Comparing the $\log L$ values, we found that the residual emission is more 
likely a point source located at the bright radio shell, as
the fit improvement of the model Radio$_{\rm ps}$ $+$ HESS$_{0.32}$ is
at 4.3$\sigma$ and 3.3$\sigma$ significance levels with respect to
the models of Radio$_{\rm disk}$ $+$ HESS$_{0.32}$ and 
Radio$_{\rm profile}$ $+$ HESS$_{0.32}$, respectively (see Table~\ref{tab:lh}).

\begin{figure}
   \centering
   \includegraphics[width=0.48\textwidth]{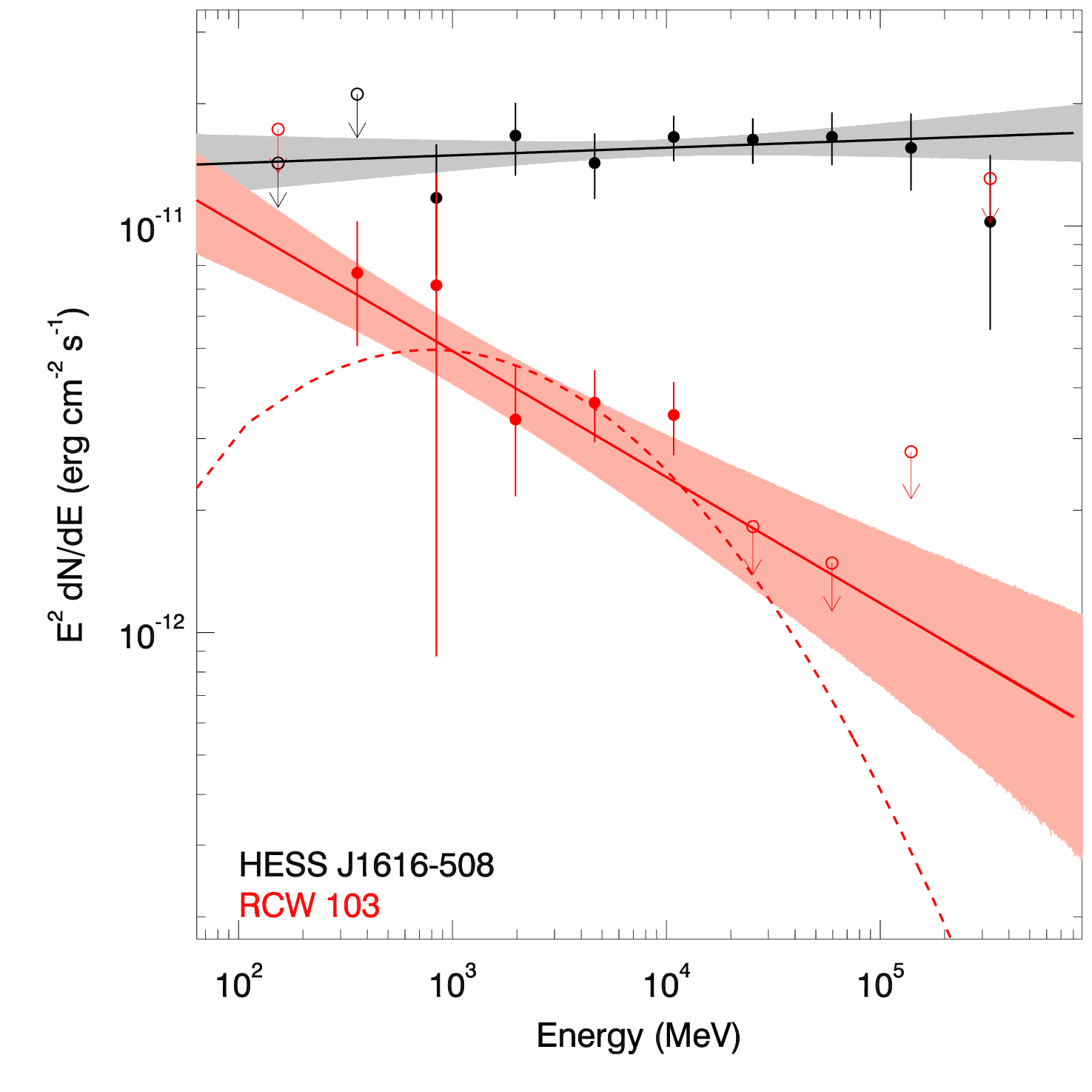}
   \caption{\gr\ spectra of \rcw\ and \hess\ and their respective spectral 
	models obtained from the likelihood analysis. For \rcw, we also
	show a Log-Parabola fit obtained in Section~\ref{sec:ss} (red dashed 
	curve).
}
   \label{fig:spectra}
\end{figure}

To be as complete as possible, we re-checked the extension of the counterpart
of \hess\ by setting Radio$_{\rm ps}$ for \rcw.
The uniform disk with a radius of 0\fdg02--0\fdg50 (in a 0\fdg02 step)
was tested again. We found that a 0\fdg38 radius (thus the HESS$_{0.38}$ model
used above) resulted in a maximum $\log L$ value and
the fit improvement of the model of Radio$_{\rm ps}$ $+$ HESS$_{0.38}$
is at a 4.9$\sigma$ significance level with respect to that of
Radio$_{\rm ps}$ $+$ HESS$_{0.32}$ (Table~\ref{tab:lh}). 
We thus considered this Radio$_{\rm ps}$ $+$ HESS$_{0.38}$ model as the final 
one and used it in the following analysis.

Based on this model, we obtained $\Gamma= 1.98\pm$0.03, 
$F_{0.3-500}\simeq (3.1\pm0.3)\times 10^{-8}$\,photon\,cm$^{-2}$\,s$^{-1}$
for the counterpart of \hess\ (TS=1233) and $\Gamma=2.31\pm$0.07, 
$F_{0.3-500}\simeq 1.1\pm0.2\times 10^{-8}$\,photon\,cm$^{-2}$\,s$^{-1}$ 
for \rcw\ (TS=165). Compared to the results previously reported 
in \citet{xing+14}, in which a point source was considered for \hess,
the emission has a higher $\Gamma$ value (or softer). The \gr\ luminosity is 
2.4$\times 10^{34}$\,erg\,s$^{-1}$ at a distance of 3.1\,kpc.

We also tested a Log-Parabola model, 
$dN/dE = N_{0}(E/E_{b})^{[-\alpha - \beta log(E/E_{b})]}$, for the emission
of RCW 103. The parameters $\alpha$ and $\beta$ were set free, and
$E_{b}$, a scale parameter, was always fixed. Different $E_{b}$ values were
tested. If we fixed it at 5\,GeV, the resulting likelihood value
indicated the fitting was better than that with the PL at a $\sim$2$\sigma$ 
significance level, where $\alpha = 2.4+/-0.6$, $\beta= 0.11+/-0.04$, and 
$F_{0.3-500} \simeq (9.5\pm 3.6)\times 10^{-9}$ photon\,cm$^{-2}$\,s$^{-1}$
(TS $\simeq 166$). The results are similar to those with the PL (note that
the curvature parameter
$\beta$ is very small), but the uncertainties are larger.

\subsection{Spectral Analysis}

We extracted the \gr\ spectra of \hess\ and \rcw\ by performing the maximum 
likelihood analysis of the LAT data in 10 evenly divided energy bands 
in logarithm from 0.1 to 500\,GeV. 
In the extraction, the spectral normalizations of the sources within 5\,degree
of the center of \hess\ were set as free parameters, while all the other 
parameters were fixed at the values obtained from the maximum 
likelihood analysis (the Radio$_{\rm ps}$ $+$ HESS$_{0.38}$ model). 
Emissions of \hess\ and \rcw\ were respectively described by a PL, 
with $\Gamma=2$.
For the obtained spectral data points, we kept those with TS$\geq$9 and
used the derived 95\% flux upper limits instead when TS$<$9 
(Table~\ref{tab:spectra}).

We also evaluated the systematic uncertainties induced by the uncertainties 
of the Galactic diffuse emission. Following the commonly-used method 
(e.g., \citealt{abdo+w51c2009,abdo+w28-2010}), we repeated the likelihood 
analysis in each energy band with the normalizations of the Galactic diffuse 
component artificially fixed to the $\pm$6\% deviation values from the 
best-fit values. This $\pm$6\% deviation represents the local departure from 
the best-fit diffuse model obtained from analyzing the source-free regions in
the Galactic plane.

The obtained spectra with both statistical and systematic uncertainties 
are shown in Figure~\ref{fig:spectra}, while the flux and 
TS values of the spectral data points are provided in Table~\ref{tab:spectra}. 
The PL spectral models obtained for the two sources from the likelihood
analysis match the spectral data points. However for \rcw, the two 
upper limits
in 17--91\,GeV (Table~\ref{tab:spectra}) are at the model-fit line, showing
a slight inconsistency. Thus, its emission may be better described with
the Log-Parabola model we obtained in Section~\ref{sec:ss}, which is also
shown in Figure~\ref{fig:spectra}.

\begin{table}
\begin{center}
\caption{Flux Measurements}
\label{tab:spectra}
\begin{tabular}{lccccc}
\hline
	Band (GeV) & $G_{\rm HESS}$ & TS & $G_{\rm RCW}$ & TS \\ \hline
0.15 (0.1--0.2) & 1.43 & 0 & 1.73 & 0 \\
0.36 (0.2--0.5) & 2.11 & 0 & 0.77$\pm$0.26 & 44 \\
0.84 (0.5--1.3) & 1.17$\pm$0.41 & 119 & 0.72$\pm$0.63 & 67 \\
1.97 (1.3--3.0) & 1.67$\pm$0.34 & 294 & 0.33$\pm$0.12 & 26 \\
4.62 (3.0--7.1) & 1.43$\pm$0.26 & 198 & 0.37$\pm$0.07 & 47 \\
10.83 (7.1--16.6) & 1.66$\pm$0.21 & 210 & 0.34$\pm$0.07 & 41 \\
25.37 (16.6--38.8) & 1.63$\pm$0.21 & 146 & 0.18 & 1 \\
59.46 (38.8--91.0) & 1.66$\pm$0.25 & 89 & 0.15 & 0 \\
139.36 (91.0--213.3) & 1.56$\pm$0.33 & 44 & 0.28 & 0 \\
326.60 (213.3--500.0) & 1.02$\pm$0.47 & 10 & 1.31 & 6 \\
\hline
\end{tabular}
\\
\footnotesize{Note: $G$ is the energy flux ($E^{2} dN/dE$) in 10$^{-11}$ erg cm$^{-2}$ s$^{-1}$.  Fluxes without uncertainties are 
the 95$\%$ upper limits.}
\end{center}
\end{table}

\subsection{Variability Check}

We checked the \gr\ emission of \rcw\ for any long-term 
variability. Following the procedure introduced in \citet{nol+12}, 
the variability index TS$_{var}$ of a 60-day binned light curve
(94 time bins) in 0.3--500 GeV was calculated, and we obtained
TS$_{var}\simeq 62.8$. Given that
a variable source would be identified at a 99\% confidence level
when TS$_{var}\geq$127.6 (for 93 degrees of freedom), the emission did not
show significant long-term variability.

\section{Discussion}

Using the latest \fermi-LAT source catalog, we have revisited the detection of
\gr\ emission of \rcw\ by analyzing the data collected for $>15$ years.
The SNR's emission is resolved at high-energy bands
(Figures~\ref{fig:tsmap} \& \ref{fig:ts5gev}), even though the nearby source 
4FGL~J1616.2$-$5054e, 
the counterpart to the VHE source \hess, has been determined to be extended.
Differing from the previous result reported in \citet{xing+14}
that the emission was extended,
we have found that it is consistent with being a PS.
It has a PL spectrum with photon index $\Gamma\simeq 2.31$ or
a Log-Parabola one with slope $\alpha\simeq 2.4$,
and the spectrum is softer than the previously reported
($\Gamma\simeq 2.0$ in \citealt{xing+14}). We note that this spectrum is
very similar to that of the SNR Kes 17, which had $\Gamma\simeq 2.39$
\citep{scl23}. 

The determined error-circle region of the PS emission coincides with
the southern limb of \rcw\ (Figure~\ref{fig:tsmap}), the brightest part of the
SNR. 
No non-thermal X-ray emission has been detected from any part of \rcw\ 
(e.g., \citealt{bsf19}), which otherwise may provide
information for the origin of the \gr\ emission. On the other hand,
multi-wavelength observations have shown different pieces of evidence for 
the interaction of the remnant with a molecular cloud (MC) in the vicinity of
the southern limb (see, e.g., \citealt{oli+99,par+06,rea+06,pin+11}; see also
\citealt{bsf19} and references therein). Considering these, the positional 
coincidence thus strongly suggest a hadronic origin for the \gr\ emission.

\begin{figure}
	\centering
   \includegraphics[width=0.48\textwidth]{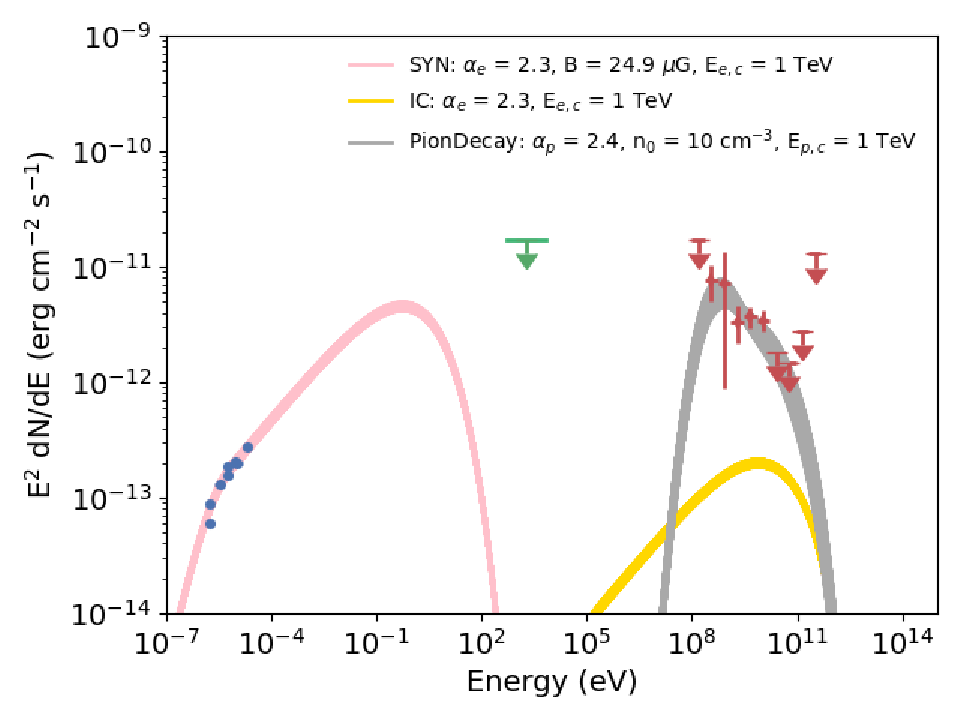}
   \caption{SED of RCW 103, including radio fluxes (blue data points), an
	X-ray flux upper limit (green arrow), and \gr\ fluxes (red data
	points). Electrons with an index$\sim 2.3$ PL energy distribution
	produce the pink and yellow spectra through synchrotron radiation
	and ICS, respectively, and protons with an index$\sim 2.4$ PL energy
	distribution produce the grey spectrum through the hadronic 
	proton-proton collisions.  }
   \label{fig:sed}
\end{figure}

We studied the broadband spectral energy distribution (SED) of the southern limb
of \rcw.
The radio flux measurements of the whole SNR were provided in 
\citet{bea66}, \citet{gs70}, \citet{sg70}, \citet{cas+80}, and \citet{dic+96}, 
and we divided them by 2 to approximate the radio fluxes of the southern part.
We analyzed archival X-ray data from one {\it Chandra} observation 
of \rcw\ (Obsid:18459), and obtained the X-ray flux
in 0.5--7\,keV from a region just outside the southern limb as the upper limit 
on non-thermal X-ray emission, where the value (unabsorbed) was 
$\simeq 1.7\times 10^{-11}$\,erg\,s$^{-1}$\,cm$^{-2}$. The radio fluxes and 
the X-ray flux upper limit, as well as the \gr\ fluxes, are shown in
Figure~\ref{fig:sed}.

To approximately fit the SED, we assumed that the particles accelerated in 
the SNR have a power-law form in energy with a high-energy cutoff:
\begin{equation}
    dN_i/dE = A_{i}E^{-\alpha_i} \mathrm{exp}(-E/E_{i,c})\ ,
\end{equation}
where $i=e/p$, the electrons or protons, $A_i$ is the normalization, and 
$\alpha_i$ and $E_{i,c}$ are the power-law index and 
the cutoff energy, respectively.
The total energy ($W_i$) is obtained from the integral of the equation.
We used the Python package {\tt naima} \citep{naima} to find reasonable
fits to the SED.

By fitting the radio data with synchrotron radiation of the electrons, we 
obtained $\alpha_e = 2.28\pm0.01$, magnetic field strength
$B=24.9\pm0.3$\,$\mu$G, and 
$W_e=3.8\pm0.1\times10^{47}$ erg, where $E_{e,c}$ was fixed at 1 TeV.
The same population of the electrons also upper-scatter the low-energy photons
in the field and produce \gr\ emission (i.e., the ICS process).
Assuming the cosmic microwave background and the interstellar radiation field 
with a temperature of 30 K and an energy density of 
0.5\,eV~cm$^{-3}$ \citep{por+06} at the source position, the produced
\gr\ spectrum from ICS is shown in Figure~\ref{fig:sed}, which is low
compared to the observed \gr\ fluxes. It should noted that in this 
fitting practice, the allowed parameter ranges are much larger than those 
reported above, mainly
because of the lack of a tight constraint in X-rays. Different combinations
of the parameters can provide adequate fits to the radio data. However,
we also noted
that none of the combinations can simultaneously fit the \gr\ data. 
Because of the following
hadronic model results (which have $\alpha_p \simeq 2.4$), we chose to
report the above parameters because of a similar electron PL index.

In order to fit the \gr\ data points with a hadronic model, in which 
\gr\ emission primarily arises from proton-proton collisions,
we obtained
$\alpha_p=2.4\pm0.1$ and $W_p=2.4^{+0.9}_{-0.5}\times10^{49}$ erg, 
where $E_{p,c}=1$~TeV was fixed and the target density was assumed to
be $n_0=10\ {\rm cm^{-3}}$ (also adopted in \citealt{xing+14}).
As shown in Figure~\ref{fig:sed}, the model spectrum can adequately describe
the \gr\ data points. In addition, the spectrum drops to low flux values
above $\sim 0.1$\,TeV, which may explain the non-detection of this SNR in 
the TeV band (e.g., \citealt{hgps}). The hadronic-model 
fitting results, in combination with the positional
coincidence of the \gr\ emission with the southern limb, 
thus point to the remnant-MC interaction as the likely origin
for the high-energy emission from \rcw.

%%\begin{table}
%%\begin{center}
%%\caption{Likelihood analysis results}
%%\label{tab:likelihood_results}
%%\begin{tabular}{llccccccc}
%%\hline
%%Source Model                             & source & $\Gamma$ & $F$ & TS \\ \hline
%%Radio$_{ps}$ $+$ HESS$_{0.38}$ & \hess\ & 1.98$\pm$0.03 & 3.1$\pm$0.3 & 1233 \\ 
%% & \rcw\  & 2.31$\pm$0.07 & 1.1$\pm$0.2 & 165 \\ 
%%\hline
%%\end{tabular}
%%\\
%%\footnotesize{Note: $F$ is the 0.3--500 GeV photon flux in 10$^{-8}$ photons cm$^{-2}$ s$^{-1}$.}
%%\end{center}
%%\end{table}

\begin{acknowledgements}
This research is supported by the Basic Research Program of Yunnan Province
(No. 202201AS070005), the National Natural Science Foundation of
China (12273033), and the Original
Innovation Program of the Chinese Academy of Sciences (E085021002).
D.Z. acknowledges the support of the science research program for graduate 
	students of Yunnan University (KC-23234629).
\end{acknowledgements}

\bibliographystyle{aasjournal}
\bibliography{rcw}

\begin{thebibliography}{}
\expandafter\ifx\csname natexlab\endcsname\relax\def\natexlab#1{#1}\fi

\bibitem[{{Abdo} {et~al.}(2009){Abdo}, {Ackermann}, {Ajello}, {Baldini},
  {Ballet}, {Barbiellini}, {Baring}, {Bastieri}, {Baughman}, {Bechtol},
  {Bellazzini}, {Berenji}, {Blandford}, {Bloom}, {Bonamente}, {Borgland},
  {Bouvier}, {Bregeon}, {Brez}, {Brigida}, {Bruel}, {Burnett}, {Buson},
  {Caliandro}, {Cameron}, {Caraveo}, {Casandjian}, {Cecchi}, {{\c C}elik},
  {Chekhtman}, {Cheung}, {Chiang}, {Ciprini}, {Claus}, {Cohen-Tanugi},
  {Cominsky}, {Conrad}, {Cutini}, {Dermer}, {de Angelis}, {de Palma}, {Digel},
  {Dormody}, {Silva}, {Drell}, {Dubois}, {Dumora}, {Farnier}, {Favuzzi},
  {Fegan}, {Focke}, {Fortin}, {Frailis}, {Fukazawa}, {Funk}, {Fusco},
  {Gargano}, {Gasparrini}, {Gehrels}, {Germani}, {Giavitto}, {Giebels},
  {Giglietto}, {Giordano}, {Glanzman}, {Godfrey}, {Grenier}, {Grondin},
  {Grove}, {Guillemot}, {Guiriec}, {Hanabata}, {Harding}, {Hayashida}, {Hays},
  {Hughes}, {Jackson}, {J{\'o}hannesson}, {Johnson}, {Johnson}, {Johnson},
  {Kamae}, {Katagiri}, {Kataoka}, {Katsuta}, {Kawai}, {Kerr}, {Kn{\"o}dlseder},
  {Kocian}, {Kuss}, {Lande}, {Latronico}, {Lemoine-Goumard}, {Longo},
  {Loparco}, {Lott}, {Lovellette}, {Lubrano}, {Makeev}, {Mazziotta}, {McEnery},
  {Meurer}, {Michelson}, {Mitthumsiri}, {Mizuno}, {Moiseev}, {Monte},
  {Monzani}, {Morselli}, {Moskalenko}, {Murgia}, {Nakamori}, {Nolan}, {Norris},
  {Nuss}, {Ohsugi}, {Okumura}, {Omodei}, {Orlando}, {Ormes}, {Paneque},
  {Parent}, {Pelassa}, {Pepe}, {Pesce-Rollins}, {Piron}, {Porter}, {Rain{\`o}},
  {Rando}, {Razzano}, {Reimer}, {Reimer}, {Reposeur}, {Ritz}, {Rodriguez},
  {Romani}, {Roth}, {Ryde}, {Sadrozinski}, {Sanchez}, {Sander}, {Saz
  Parkinson}, {Scargle}, {Schalk}, {Sgr{\`o}}, {Siskind}, {Smith}, {Smith},
  {Spandre}, {Spinelli}, {Strickman}, {Suson}, {Tajima}, {Takahashi},
  {Takahashi}, {Tanaka}, {Thayer}, {Thayer}, {Thompson}, {Tibaldo}, {Tibolla},
  {Torres}, {Tosti}, {Tramacere}, {Uchiyama}, {Usher}, {Vasileiou}, {Venter},
  {Vilchez}, {Vitale}, {Waite}, {Wang}, {Winer}, {Wood}, {Yamazaki}, {Ylinen},
  \& {Ziegler}}]{abdo+w51c2009}
{Abdo}, A.~A., {Ackermann}, M., {Ajello}, M., {et~al.} 2009, \apjl, 706, L1

\bibitem[{{Abdo} {et~al.}(2010){Abdo}, {Ackermann}, {Ajello}, {Allafort},
  {Baldini}, {Ballet}, {Barbiellini}, {Bastieri}, {Bechtol}, {Bellazzini},
  {Berenji}, {Blandford}, {Bloom}, {Bonamente}, {Borgland}, {Bouvier},
  {Brandt}, {Bregeon}, {Brigida}, {Bruel}, {Buehler}, {Buson}, {Caliandro},
  {Cameron}, {Caraveo}, {Carrigan}, {Casandjian}, {Cecchi}, {{\c C}elik},
  {Chekhtman}, {Chiang}, {Ciprini}, {Claus}, {Cohen-Tanugi}, {Conrad},
  {Dermer}, {de Palma}, {Silva}, {Drell}, {Dubois}, {Dumora}, {Farnier},
  {Favuzzi}, {Fegan}, {Fukazawa}, {Fukui}, {Funk}, {Fusco}, {Gargano},
  {Gehrels}, {Germani}, {Giglietto}, {Giordano}, {Glanzman}, {Godfrey},
  {Grenier}, {Grove}, {Guiriec}, {Hadasch}, {Hanabata}, {Harding}, {Hays},
  {Horan}, {Hughes}, {J{\'o}hannesson}, {Johnson}, {Johnson}, {Kamae},
  {Katagiri}, {Kataoka}, {Kn{\"o}dlseder}, {Kuss}, {Lande}, {Latronico}, {Lee},
  {Lemoine-Goumard}, {Llena Garde}, {Longo}, {Loparco}, {Lovellette},
  {Lubrano}, {Makeev}, {Mazziotta}, {Michelson}, {Mitthumsiri}, {Mizuno},
  {Moiseev}, {Monte}, {Monzani}, {Morselli}, {Moskalenko}, {Murgia},
  {Nakamori}, {Nolan}, {Norris}, {Nuss}, {Ohno}, {Ohsugi}, {Omodei}, {Orlando},
  {Ormes}, {Ozaki}, {Panetta}, {Parent}, {Pelassa}, {Pepe}, {Pesce-Rollins},
  {Piron}, {Porter}, {Rain{\`o}}, {Rando}, {Razzano}, {Reimer}, {Reimer},
  {Reposeur}, {Rodriguez}, {Roth}, {Sadrozinski}, {Sander}, {Saz Parkinson},
  {Sgr{\`o}}, {Siskind}, {Smith}, {Smith}, {Spandre}, {Spinelli}, {Strickman},
  {Suson}, {Tajima}, {Takahashi}, {Takahashi}, {Tanaka}, {Thayer}, {Thayer},
  {Thompson}, {Tibaldo}, {Tibolla}, {Torres}, {Tosti}, {Uchiyama}, {Uehara},
  {Usher}, {Vasileiou}, {Vilchez}, {Vitale}, {Waite}, {Wang}, {Winer}, {Wood},
  {Yamamoto}, {Yamazaki}, {Yang}, {Ylinen}, \& {Ziegler}}]{abdo+w28-2010}
---. 2010, \apj, 718, 348

\bibitem[{{Abdollahi} {et~al.}(2020){Abdollahi}, {Acero}, {Ackermann},
  {Ajello}, {Atwood}, {Axelsson}, {Baldini}, {Ballet}, {Barbiellini},
  {Bastieri}, {Becerra Gonzalez}, {Bellazzini}, {Berretta}, {Bissaldi},
  {Blandford}, {Bloom}, {Bonino}, {Bottacini}, {Brandt}, {Bregeon}, {Bruel},
  {Buehler}, {Burnett}, {Buson}, {Cameron}, {Caputo}, {Caraveo}, {Casandjian},
  {Castro}, {Cavazzuti}, {Charles}, {Chaty}, {Chen}, {Cheung}, {Chiaro},
  {Ciprini}, {Cohen-Tanugi}, {Cominsky}, {Coronado-Bl{\'a}zquez}, {Costantin},
  {Cuoco}, {Cutini}, {D'Ammando}, {DeKlotz}, {de la Torre Luque}, {de Palma},
  {Desai}, {Digel}, {Di Lalla}, {Di Mauro}, {Di Venere}, {Dom{\'\i}nguez},
  {Dumora}, {Fana Dirirsa}, {Fegan}, {Ferrara}, {Franckowiak}, {Fukazawa},
  {Funk}, {Fusco}, {Gargano}, {Gasparrini}, {Giglietto}, {Giommi}, {Giordano},
  {Giroletti}, {Glanzman}, {Green}, {Grenier}, {Griffin}, {Grondin}, {Grove},
  {Guiriec}, {Harding}, {Hayashi}, {Hays}, {Hewitt}, {Horan},
  {J{\'o}hannesson}, {Johnson}, {Kamae}, {Kerr}, {Kocevski}, {Kovac'evic'},
  {Kuss}, {Landriu}, {Larsson}, {Latronico}, {Lemoine-Goumard}, {Li},
  {Liodakis}, {Longo}, {Loparco}, {Lott}, {Lovellette}, {Lubrano}, {Madejski},
  {Maldera}, {Malyshev}, {Manfreda}, {Marchesini}, {Marcotulli},
  {Mart{\'\i}-Devesa}, {Martin}, {Massaro}, {Mazziotta}, {McEnery}, {Mereu},
  {Meyer}, {Michelson}, {Mirabal}, {Mizuno}, {Monzani}, {Morselli},
  {Moskalenko}, {Negro}, {Nuss}, {Ojha}, {Omodei}, {Orienti}, {Orlando},
  {Ormes}, {Palatiello}, {Paliya}, {Paneque}, {Pei}, {Pe{\~n}a-Herazo},
  {Perkins}, {Persic}, {Pesce-Rollins}, {Petrosian}, {Petrov}, {Piron}, {Poon},
  {Porter}, {Principe}, {Rain{\`o}}, {Rando}, {Razzano}, {Razzaque}, {Reimer},
  {Reimer}, {Remy}, {Reposeur}, {Romani}, {Saz Parkinson}, {Schinzel},
  {Serini}, {Sgr{\`o}}, {Siskind}, {Smith}, {Spandre}, {Spinelli}, {Strong},
  {Suson}, {Tajima}, {Takahashi}, {Tak}, {Thayer}, {Thompson}, {Tibaldo},
  {Torres}, {Torresi}, {Valverde}, {Van Klaveren}, {van Zyl}, {Wood},
  {Yassine}, \& {Zaharijas}}]{4fgl}
{Abdollahi}, S., {Acero}, F., {Ackermann}, M., {et~al.} 2020, \apjs, 247, 33

\bibitem[{{Aharonian} {et~al.}(2005){Aharonian}, {Akhperjanian}, {Aye},
  {Bazer-Bachi}, {Beilicke}, {Benbow}, {Berge}, {Berghaus}, {Bernl{\"o}hr},
  {Boisson}, {Bolz}, {Borgmeier}, {Braun}, {Breitling}, {Brown}, {Gordo},
  {Chadwick}, {Chounet}, {Cornils}, {Costamante}, {Degrange},
  {Djannati-Ata{\"\i}}, {Drury}, {Dubus}, {Ergin}, {Espigat}, {Feinstein},
  {Fleury}, {Fontaine}, {Funk}, {Gallant}, {Giebels}, {Gillessen}, {Goret},
  {Hadjichristidis}, {Hauser}, {Heinzelmann}, {Henri}, {Hermann}, {Hinton},
  {Hofmann}, {Holleran}, {Horns}, {de Jager}, {Jung}, {Kh{\'e}lifi}, {Komin},
  {Konopelko}, {Latham}, {Le Gallou}, {Lemi{\`e}re}, {Lemoine}, {Leroy},
  {Lohse}, {Marcowith}, {Masterson}, {McComb}, {de Naurois}, {Nolan},
  {Noutsos}, {Orford}, {Osborne}, {Ouchrif}, {Panter}, {Pelletier}, {Pita},
  {P{\"u}hlhofer}, {Punch}, {Raubenheimer}, {Raue}, {Raux}, {Rayner},
  {Redondo}, {Reimer}, {Reimer}, {Ripken}, {Rob}, {Rolland}, {Rowell},
  {Sahakian}, {Saug{\'e}}, {Schlenker}, {Schlickeiser}, {Schuster}, {Schwanke},
  {Siewert}, {Sol}, {Steenkamp}, {Stegmann}, {Tavernet}, {Terrier},
  {Th{\'e}oret}, {Tluczykont}, {van der Walt}, {Vasileiadis}, {Venter},
  {Vincent}, {Visser}, {V{\"o}lk}, \& {Wagner}}]{hess05}
{Aharonian}, F., {Akhperjanian}, A.~G., {Aye}, K.~M., {et~al.} 2005, Science,
  307, 1938

\bibitem[{{Aharonian} {et~al.}(2006){Aharonian}, {Akhperjanian}, {Bazer-Bachi},
  {Beilicke}, {Benbow}, {Berge}, {Bernl{\"o}hr}, {Boisson}, {Bolz}, {Borrel},
  {Braun}, {Breitling}, {Brown}, {Chadwick}, {Chounet}, {Cornils},
  {Costamante}, {Degrange}, {Dickinson}, {Djannati-Ata{\"\i}}, {Drury},
  {Dubus}, {Emmanoulopoulos}, {Espigat}, {Feinstein}, {Fontaine}, {Fuchs},
  {Funk}, {Gallant}, {Giebels}, {Gillessen}, {Glicenstein}, {Goret},
  {Hadjichristidis}, {Hauser}, {Heinzelmann}, {Henri}, {Hermann}, {Hinton},
  {Hofmann}, {Holleran}, {Horns}, {Jacholkowska}, {de Jager}, {Kh{\'e}lifi},
  {Komin}, {Konopelko}, {Latham}, {Le Gallou}, {Lemi{\`e}re},
  {Lemoine-Goumard}, {Leroy}, {Lohse}, {Martin}, {Martineau-Huynh},
  {Marcowith}, {Masterson}, {McComb}, {de Naurois}, {Nolan}, {Noutsos},
  {Orford}, {Osborne}, {Ouchrif}, {Panter}, {Pelletier}, {Pita},
  {P{\"u}hlhofer}, {Punch}, {Raubenheimer}, {Raue}, {Raux}, {Rayner}, {Reimer},
  {Reimer}, {Ripken}, {Rob}, {Rolland}, {Rowell}, {Sahakian}, {Saug{\'e}},
  {Schlenker}, {Schlickeiser}, {Schuster}, {Schwanke}, {Siewert}, {Sol},
  {Spangler}, {Steenkamp}, {Stegmann}, {Tavernet}, {Terrier}, {Th{\'e}oret},
  {Tluczykont}, {Vasileiadis}, {Venter}, {Vincent}, {V{\"o}lk}, \&
  {Wagner}}]{hess06}
{Aharonian}, F., {Akhperjanian}, A.~G., {Bazer-Bachi}, A.~R., {et~al.} 2006,
  \apj, 636, 777

\bibitem[{{Ballet} {et~al.}(2023){Ballet}, {Bruel}, {Burnett}, {Lott}, \& {The
  Fermi-LAT collaboration}}]{4fgl-dr4}
{Ballet}, J., {Bruel}, P., {Burnett}, T.~H., {Lott}, B., \& {The Fermi-LAT
  collaboration}. 2023, arXiv e-prints, arXiv:2307.12546

\bibitem[{{Beard}(1966)}]{bea66}
{Beard}, M. 1966, Australian Journal of Physics, 19, 141

\bibitem[{{Braun} {et~al.}(2019){Braun}, {Safi-Harb}, \& {Fryer}}]{bsf19}
{Braun}, C., {Safi-Harb}, S., \& {Fryer}, C.~L. 2019, \mnras, 489, 4444

\bibitem[{{Carter} {et~al.}(1997){Carter}, {Dickel}, \& {Bomans}}]{cdb97}
{Carter}, L.~M., {Dickel}, J.~R., \& {Bomans}, D.~J. 1997, \pasp, 109, 990

\bibitem[{{Caswell} {et~al.}(1980){Caswell}, {Haynes}, {Milne}, \&
  {Wellington}}]{cas+80}
{Caswell}, J.~L., {Haynes}, R.~F., {Milne}, D.~K., \& {Wellington}, K.~J. 1980,
  \mnras, 190, 881

\bibitem[{{D'A{\`\i}} {et~al.}(2016){D'A{\`\i}}, {Evans}, {Burrows}, {Kuin},
  {Kann}, {Campana}, {Maselli}, {Romano}, {Cusumano}, {La Parola}, {Barthelmy},
  {Beardmore}, {Cenko}, {De Pasquale}, {Gehrels}, {Greiner}, {Kennea}, {Klose},
  {Melandri}, {Nousek}, {Osborne}, {Palmer}, {Sbarufatti}, {Schady}, {Siegel},
  {Tagliaferri}, {Yates}, \& {Zane}}]{dai+16}
{D'A{\`\i}}, A., {Evans}, P.~A., {Burrows}, D.~N., {et~al.} 2016, \mnras, 463,
  2394

\bibitem[{{Dickel} {et~al.}(1996){Dickel}, {Green}, {Ye}, \& {Milne}}]{dic+96}
{Dickel}, J.~R., {Green}, A., {Ye}, T., \& {Milne}, D.~K. 1996, \aj, 111, 340

\bibitem[{{Drury}(1983)}]{dru83}
{Drury}, L.~O. 1983, Reports on Progress in Physics, 46, 973

\bibitem[{{Goss} \& {Shaver}(1970)}]{gs70}
{Goss}, W.~M., \& {Shaver}, P.~A. 1970, Australian Journal of Physics
  Astrophysical Supplement, 14, 1

\bibitem[{{H.~E.~S.~S. Collaboration} {et~al.}(2018){H.~E.~S.~S.
  Collaboration}, {Abdalla}, {Abramowski}, {Aharonian}, {Ait Benkhali},
  {Ang{\"u}ner}, {Arakawa}, {Arrieta}, {Aubert}, {Backes}, {Balzer}, {Barnard},
  {Becherini}, {Becker Tjus}, {Berge}, {Bernhard}, {Bernl{\"o}hr}, {Blackwell},
  {B{\"o}ttcher}, {Boisson}, {Bolmont}, {Bonnefoy}, {Bordas}, {Bregeon},
  {Brun}, {Brun}, {Bryan}, {B{\"u}chele}, {Bulik}, {Capasso}, {Carrigan},
  {Caroff}, {Carosi}, {Casanova}, {Cerruti}, {Chakraborty}, {Chaves}, {Chen},
  {Chevalier}, {Colafrancesco}, {Condon}, {Conrad}, {Davids}, {Decock}, {Deil},
  {Devin}, {deWilt}, {Dirson}, {Djannati-Ata{\"\i}}, {Domainko}, {Donath},
  {Drury}, {Dutson}, {Dyks}, {Edwards}, {Egberts}, {Eger}, {Emery},
  {Ernenwein}, {Eschbach}, {Farnier}, {Fegan}, {Fernandes}, {Fiasson},
  {Fontaine}, {F{\"o}rster}, {Funk}, {F{\"u}{\ss}ling}, {Gabici}, {Gallant},
  {Garrigoux}, {Gast}, {Gat{\'e}}, {Giavitto}, {Giebels}, {Glawion},
  {Glicenstein}, {Gottschall}, {Grondin}, {Hahn}, {Haupt}, {Hawkes},
  {Heinzelmann}, {Henri}, {Hermann}, {Hinton}, {Hofmann}, {Hoischen}, {Holch},
  {Holler}, {Horns}, {Ivascenko}, {Iwasaki}, {Jacholkowska}, {Jamrozy},
  {Jankowsky}, {Jankowsky}, {Jingo}, {Jouvin}, {Jung-Richardt}, {Kastendieck},
  {Katarzy{\'n}ski}, {Katsuragawa}, {Katz}, {Kerszberg}, {Khangulyan},
  {Kh{\'e}lifi}, {King}, {Klepser}, {Klochkov}, {Klu{\'z}niak}, {Komin},
  {Kosack}, {Krakau}, {Kraus}, {Kr{\"u}ger}, {Laffon}, {Lamanna}, {Lau},
  {Lees}, {Lefaucheur}, {Lemi{\`e}re}, {Lemoine-Goumard}, {Lenain}, {Leser},
  {Lohse}, {Lorentz}, {Liu}, {L{\'o}pez-Coto}, {Lypova}, {Marandon},
  {Malyshev}, {Marcowith}, {Mariaud}, {Marx}, {Maurin}, {Maxted}, {Mayer},
  {Meintjes}, {Meyer}, {Mitchell}, {Moderski}, {Mohamed}, {Mohrmann},
  {Mor{\r{a}}}, {Moulin}, {Murach}, {Nakashima}, {de Naurois}, {Ndiyavala},
  {Niederwanger}, {Niemiec}, {Oakes}, {O'Brien}, {Odaka}, {Ohm}, {Ostrowski},
  {Oya}, {Padovani}, {Panter}, {Parsons}, {Paz Arribas}, {Pekeur}, {Pelletier},
  {Perennes}, {Petrucci}, {Peyaud}, {Piel}, {Pita}, {Poireau}, {Poon},
  {Prokhorov}, {Prokoph}, {P{\"u}hlhofer}, {Punch}, {Quirrenbach}, {Raab},
  {Rauth}, {Reimer}, {Reimer}, {Renaud}, {de los Reyes}, {Rieger}, {Rinchiuso},
  {Romoli}, {Rowell}, {Rudak}, {Rulten}, {Safi-Harb}, {Sahakian}, {Saito},
  {Sanchez}, {Santangelo}, {Sasaki}, {Schandri}, {Schlickeiser},
  {Sch{\"u}ssler}, {Schulz}, {Schwanke}, {Schwemmer}, {Seglar-Arroyo},
  {Settimo}, {Seyffert}, {Shafi}, {Shilon}, {Shiningayamwe}, {Simoni}, {Sol},
  {Spanier}, {Spir-Jacob}, {Stawarz}, {Steenkamp}, {Stegmann}, {Steppa},
  {Sushch}, {Takahashi}, {Tavernet}, {Tavernier}, {Taylor}, {Terrier},
  {Tibaldo}, {Tiziani}, {Tluczykont}, {Trichard}, {Tsirou}, {Tsuji}, {Tuffs},
  {Uchiyama}, {van der Walt}, {van Eldik}, {van Rensburg}, {van Soelen},
  {Vasileiadis}, {Veh}, {Venter}, {Viana}, {Vincent}, {Vink}, {Voisin},
  {V{\"o}lk}, {Vuillaume}, {Wadiasingh}, {Wagner}, {Wagner}, {Wagner}, {White},
  {Wierzcholska}, {Willmann}, {W{\"o}rnlein}, {Wouters}, {Yang}, {Zaborov},
  {Zacharias}, {Zanin}, {Zdziarski}, {Zech}, {Zefi}, {Ziegler}, {Zorn}, \&
  {{\.Z}ywucka}}]{hgps}
{H.~E.~S.~S. Collaboration}, {Abdalla}, H., {Abramowski}, A., {et~al.} 2018,
  \aap, 612, A1

\bibitem[{{Leahy} {et~al.}(2020){Leahy}, {Ranasinghe}, \& {Gelowitz}}]{lrg20}
{Leahy}, D.~A., {Ranasinghe}, S., \& {Gelowitz}, M. 2020, \apjs, 248, 16

\bibitem[{{Nolan} {et~al.}(2012){Nolan}, {Abdo}, {Ackermann}, {Ajello},
  {Allafort}, {Antolini}, {Atwood}, {Axelsson}, {Baldini}, {Ballet}, \&
  et~al.}]{nol+12}
{Nolan}, P.~L., {Abdo}, A.~A., {Ackermann}, M., {et~al.} 2012, \apjs, 199, 31

\bibitem[{{Oliva} {et~al.}(1999){Oliva}, {Moorwood}, {Drapatz}, {Lutz}, \&
  {Sturm}}]{oli+99}
{Oliva}, E., {Moorwood}, A.~F.~M., {Drapatz}, S., {Lutz}, D., \& {Sturm}, E.
  1999, \aap, 343, 943

\bibitem[{{Paron} {et~al.}(2006){Paron}, {Reynoso}, {Purcell}, {Dubner}, \&
  {Green}}]{par+06}
{Paron}, S.~A., {Reynoso}, E.~M., {Purcell}, C., {Dubner}, G.~M., \& {Green},
  A. 2006, \pasa, 23, 69

\bibitem[{{Pinheiro Gon{\c{c}}alves} {et~al.}(2011){Pinheiro Gon{\c{c}}alves},
  {Noriega-Crespo}, {Paladini}, {Martin}, \& {Carey}}]{pin+11}
{Pinheiro Gon{\c{c}}alves}, D., {Noriega-Crespo}, A., {Paladini}, R., {Martin},
  P.~G., \& {Carey}, S.~J. 2011, \aj, 142, 47

\bibitem[{{Porter} {et~al.}(2006){Porter}, {Moskalenko}, \& {Strong}}]{por+06}
{Porter}, T.~A., {Moskalenko}, I.~V., \& {Strong}, A.~W. 2006, \apjl, 648, L29

\bibitem[{{Rea} {et~al.}(2016){Rea}, {Borghese}, {Esposito}, {Coti Zelati},
  {Bachetti}, {Israel}, \& {De Luca}}]{rbe+16}
{Rea}, N., {Borghese}, A., {Esposito}, P., {et~al.} 2016, \apjl, 828, L13

\bibitem[{{Reach} {et~al.}(2006){Reach}, {Rho}, {Tappe}, {Pannuti}, {Brogan},
  {Churchwell}, {Meade}, {Babler}, {Indebetouw}, \& {Whitney}}]{rea+06}
{Reach}, W.~T., {Rho}, J., {Tappe}, A., {et~al.} 2006, \aj, 131, 1479

\bibitem[{{Reynoso} {et~al.}(2004){Reynoso}, {Green}, {Johnston}, {Goss},
  {Dubner}, \& {Giacani}}]{rgj+04}
{Reynoso}, E.~M., {Green}, A.~J., {Johnston}, S., {et~al.} 2004, \pasa, 21, 82

\bibitem[{{Rodgers} {et~al.}(1960){Rodgers}, {Campbell}, \& {Whiteoak}}]{rcw60}
{Rodgers}, A.~W., {Campbell}, C.~T., \& {Whiteoak}, J.~B. 1960, \mnras, 121,
  103

\bibitem[{{Shaver} \& {Goss}(1970)}]{sg70}
{Shaver}, P.~A., \& {Goss}, W.~M. 1970, Australian Journal of Physics
  Astrophysical Supplement, 14, 77

\bibitem[{{Sup{\'a}n} {et~al.}(2023){Sup{\'a}n}, {Castelletti}, \&
  {Lemi{\`e}re}}]{scl23}
{Sup{\'a}n}, L., {Castelletti}, G., \& {Lemi{\`e}re}, A. 2023, \aap, 679, A22

\bibitem[{{Tuohy} \& {Garmire}(1980)}]{tg80}
{Tuohy}, I., \& {Garmire}, G. 1980, \apjl, 239, L107

\bibitem[{{Xing} {et~al.}(2014){Xing}, {Wang}, {Zhang}, \& {Chen}}]{xing+14}
{Xing}, Y., {Wang}, Z., {Zhang}, X., \& {Chen}, Y. 2014, \apj, 781, 64

\bibitem[{{Zabalza}(2015)}]{naima}
{Zabalza}, V. 2015, in International Cosmic Ray Conference, Vol.~34, 34th
  International Cosmic Ray Conference (ICRC2015), 922

\bibitem[{{Zeng} {et~al.}(2019){Zeng}, {Xin}, \& {Liu}}]{zyl19}
{Zeng}, H., {Xin}, Y., \& {Liu}, S. 2019, \apj, 874, 50

\end{thebibliography}

\end{document}